\documentclass[10pt,final]{iopart}

\usepackage[english]{babel}
\usepackage{graphicx}
\usepackage{iopams,cite}  
\usepackage{color}

\newcommand{\pderv}[2]{\frac{\partial #1}{\partial #2}}
\newcommand{\pdervi}[2]{\partial #1\slash\partial #2}
\newcommand{\derv}[2]{\frac{{\rm d} #1}{{\rm d} #2}}
\newcommand{\dervi}[2]{{\rm d} #1/{\rm d} #2}
\newcommand{\rem}[1]{}
\newcommand{\ind}[1]{{\mathrm{#1}}}

\begin{document}
\title[Interpretation of quasi-periodic frequency sweeping in electron cyclotron emission...]
{Interpretation of quasi-periodic frequency sweeping in electron cyclotron emission of nonequilibrium mirror-confined plasma sustained by high-power microwaves}
\author{A G Shalashov, E D Gospodchikov, M E Viktorov} 
\address{Institute of Applied Physics of Russian Academy of Sciences, 46 Ulyanov Str., 603950, Nizhny Novgorod, Russia}
\ead{ags@appl.sci-nnov.ru}

\begin{abstract}
In [Viktorov M E \textit{et al. }2016 \textit{Eur. Phys. Lett.} \textbf{116} 55001] we reported on chirping frequency patterns that have been observed in the electron cyclotron emission from strongly nonequilibrium plasma confined in a table-top mirror magnetic trap. These patterns were interpreted qualitatively as the first experimental evidence for so-called ``holes and clumps'' mechanism acting at such high frequencies (1--10 GHz). In the present paper, we prove this statement on a more rigorous basis, considering the formation of nonlinear phase space structures in proximity of the wave-particle resonances within the framework of the Berk--Breizman theory adopted for a kinetically unstable mirror-confined plasma. 

\end{abstract}

\pacs{52.72.+v, 52.35.-g}
\submitto{\PPCF}
\ioptwocol
\maketitle

\section{Introduction}

From a theoretical point of view, there are two limiting cases that describe resonant wave-particle interaction in plasmas. One is the quasilinear theory, a perturbative approach that involves many overlapped wave-particle resonances as a basis for a diffusive particle transport in phase space \cite{ql1,ql2}. Overlapping is provided either by a broad spectrum of excited electromagnetic waves or by a spatial inhomogeneity of resonant conditions that results in the stochastic phase of particles during repeated passing through the resonance region \cite{Tim,Tok}. The latter concept is widely used for a description of high-frequency heating of fusion plasmas by monochromatic waves \cite{Gir1,Gir2}. 

Another universal mechanism is based on a formation of separate  phase space structures in proximity of nonlinear wave-particle resonances of a kinetically unstable  plasma mode. Here the global transport is suppressed as the resonances do not overlap. As wave amplitude grows, most of the particles respond adiabatically, and only a small  group of resonant particles mix and cause local flattening of the distribution function in phase space within or near the separatrices formed by the wave. In a collisionless limit, such regimes were  proposed as a possible mechanism of generation of narrowband chorus emissions in the Earth's magnetosphere  \cite{trakh4,trakh5,Demekhov}. Similar physics was considered in \cite{berk96,berk97, brei97,berk99,nyq14} in the case of essential dissipation, which is more relevant to most laboratory experiments. 
When the linear dissipation from a background plasma is present, the saturated plateau state becomes unstable and the mode tends to grow explosively. That results in the formation and subsequent evolution of long-living  {(compared to the linear growth time)} structures in the particle distribution, so-called ``holes and clumps'' (depletion and excess of resonant particles). Their subsequent convective motion in phase space is synchronized to the change in wave frequency, thus leading to complex chirping patterns in a dynamical spectrum of unstable waves. This model, commonly referred to as the Berk--Breizman theory, has been used to study the Alfven wave turbulence driven by {high-energy ions} in toroidal magnetic traps \cite{brei97,lil10,nyq12,nyq13,pinches04,lil09,les10,les13,kosuga12,hole14}; as well as some other MHD processes, such as fishbone  \cite{brei97} and ion-acoustic \cite{les14} instabilities, have been considered. 

\begin{figure}[bt]
\centering
\includegraphics[width=83mm]{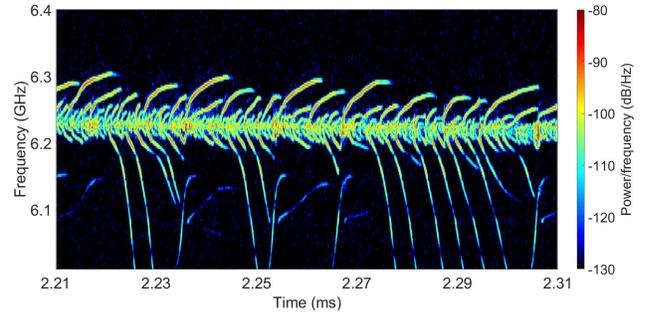}
\caption{Dynamic spectrum of enhanced microwave emission during the decay stage of nitrogen plasma after ECR breakdown in the SMIS-37 open trap facility \cite{viktorov_2016_EPL}. 
The ECR heating is operating during 0--1\,{ms} (not shown in the plot). }
\label{fig1}
\end{figure}

This paper was motivated by the first laboratory observations of potentially the same mechanism acting in a much higher electron cyclotron (EC) frequency domain reported in \cite{viktorov_2016_EPL}. We study the electron cyclotron kinetic instabilities of plasma with fast electrons sustained by high-power microwaves (with maximum intensity about 10 kW/cm$^2$ at 37.5 GHz) and confined in the table-top mirror magnetic trap. Resonant heating results in the plasma with at least two electron components, one of which, more dense and cold, determines the dispersion properties of the high-frequency waves, and the second, a small group of energetic electrons with a highly anisotropic velocity distribution, is responsible for the excitation  of unstable waves. On a phase of discharge decay after the microwave heating is switched-off, we find the chirping frequency patterns in the plasma emission in the EC frequency range (1--10 GHz) that are very similar to those predicted by the Berk--Breizman theory, see \fref{fig1}. In the present work, we adopt this theory to our case and demonstrate that separate nonlinear wave-particle resonances are indeed possible in the described experiment in spite of the huge variation of the magnetic field typical of a mirror magnetic trap.

The latter was a rather unexpected conclusion. 
During more than a decade of investigations, we observed many types of plasma EC instabilities driven by high-power microwaves with very different time and frequency patterns from a stationary generation of enhanced line emission to quasi-periodic or even stochastic burst regimes. Most of the instabilities have been adequately interpreted within the cyclotron maser paradigm, which is essentially based on the self-consistent quasilinear theory, see review \cite{shalash_2017_pop} and references therein.  In this context, a laboratory modeling of wave-particle interaction processes have much in common with similar processes occurring in the magnetosphere of the Earth \cite{trakh_book1,trakh_book2,Bingham_2013,VanCompernolle_2016,An_2016}, other planets\cite{Menietti_2012_Jupiter_Saturn}, and stars \cite{Trigilio_2011_stars}. Unlike all other EC instabilities detected at our setup, the chirping electromagnetic bursts were never accompanied with synchronous enhanced precipitations of fast electrons through the trap ends. This is an additional evidence that the quasilinear model may not be applicable to this particular case.

\def\oc{\omega_\ind{c}}
\def\occ{\omega_\ind{c0}}
\def\oca{\langle \omega_\ind{c}\rangle}
\def\ocar{\langle \tilde\omega_\ind{c}\rangle}
\def\ob{\omega_\ind{b}}
\def\o{\omega_0}

To simplify further reading, let us introduce a specific language common for both discussed mechanisms of plasma-wave interaction. 
In a mirror configuration, electrons are involved in two oscillatory motions. One is cyclotron gyration with a frequency $\oc$, the other is bouncing along the magnetic field line with a frequency {$\ob$.  In} terms of the Berk--Breizman theory, the resonance with some plasma mode at a frequency $\o$ is possible when
\begin{equation}\label{res}
n\ob+s\oca=\o,
\end{equation}
where $n, s$ are integers, and $\oca$ is a cyclotron frequency averaged over the bounce oscillations. {Noting that $\ob/\oca\sim \rho_\ind{c}/l$, where $\rho_\ind{c}$ is the Larmor radius and $l$ is the magnetic field inhomogeneity scale, one may recognize that for the majority of electrons the bounce oscillations are much slower than the cyclotron gyration.} Then, considering the electron cyclotron {domain} $s\oca\sim\o$, resonances \eref{res} for different $n$ overlap resulting in the global quasilinear diffusion.  However, under specific circumstances one or few neighboring $n$-resonances may be essentially stronger than others (characterizing by a more efficient coupling between particles and waves or by a larger stock of free energy). This provides favorable conditions for the ``holes and clumps'' formation and evolution. In the discussed experiment, this is possible due to a combination of two factors:
\begin{itemize}
	\item $\ob/\oca$ grows with the electron energy, so this quantity becomes moderately low for very fast  electrons, e.g., $\ob/\oca{\sim\rho_\ind{c}/l}\sim1/30$ for the case studied in this paper; 
	\item plasma becomes so rarefied that collisional stochastization of the electron phase is negligible, $\nu_\ind{coll}\ll\ob$, and resonances \eref{res} corresponding to moderately large $n$, e.g., $n\sim 30$, do not overlap.  
\end{itemize}
Thus, the Berk--Breizman scenario may be realized for sub-relativistic and relativistic electrons at the plasma decay stage when the background plasma density drops by few orders of magnitude compared to its values during the electron acceleration stage. 

Hereafter, this picture is elaborated in a more strict way.
We adopt the general theory of nonlinear wave-particle resonances formulated in \cite{brei97} to a particular case of magnetized electrons bouncing in a mirror magnetic configuration. {Studied effects are pronounced when $\rho_\ind{c}/l$ is not drastically small. However we assume a standard ordering of the drift theory, 
$\tau_\ind{c}\ll \tau_\ind{b}\ll\tau_\ind{dr}$,
where $\tau_\ind{c}=2\pi/\oca$, $\tau_\ind{b}=2\pi/\ob\sim (l/\rho_\ind{c})\,\tau_\ind{c}$ and $\tau_\ind{dr}\sim(l/\rho_\ind{c})^2\, \tau_\ind{c}$ are, correspondingly,  the characteristic times of gyromotion, bounce oscillations and drift across the magnetic field. Drifts across the magnetic field may be neglected for the fast time-scales considered in this paper. We also consider axisymetric magnetic configurations with a conserving angular momentum of particles, then drift orbits lay on closed flux surfaces and can not contribute to additional losses even on longer times including many bounce oscillations.}

\section{Basic equations} 
\subsection{Non-perturbed Hamiltonian}

Omitting drifts across the magnetic field, one may consider two degrees of freedom of a single electron -- the phase of cyclotron gyration $\varphi$ and the coordinate $z$ along the field line; $z=0$ stands for the minimum of the magnetic field strength. Canonically conjugated momenta are, correspondingly, the transverse adiabatic invariant $I_\perp=p_\perp^2/(2m\oc)$ and the longitudinal momentum $p_z$. Here ``transverse'' and ``longitudinal'' are related to the direction of the external magnetic field, $m$ is the electron rest mass, and $\oc=eB(z)/(mc)$ is the non-relativistic electron cyclotron frequency which depends on $z$. In the absence of unstable waves, a nonperturbed electron motion in non-relativistic limit may be described with the following  Hamiltonian with so-called Yushmanov potential\cite{yush}
\begin{equation*}
H_0=p_z^2/(2m)+I_\perp \oc(z).
\end{equation*}
Next, using a standard technique we may switch to the action-angle variables. Namely, introducing the longitudinal action,
\begin{equation*}
I_{||}=\frac{1}{2\pi}\oint \sqrt{2m\left(H_0-I_\perp\oc(z)\right)}\,\rmd z,
\end{equation*}
and expressing the Hamiltonian as a function of actions, $H_0(I_\perp,I_{||})$, we define the generating function $S$ for the desired canonical transformation as
\begin{equation*}
S(I_\perp,I_{||},\varphi,z)=I_\perp\varphi+\int^z \sqrt{2m\left(H_0-I_\perp\oc(z')\right)}\,\rmd z'.
\end{equation*}
The corresponding angles vary as
\begin{equation}\label{eq1}
\eqalign{\dervi{\xi_\perp}{t}=\pdervi{H_0}{I_\perp}=\frac\ob{2\pi}\oint\oc(z)\frac{\rmd z}{\dot z}=\oca,\cr
\dervi{\xi_{||}}{t}=\pdervi{H_0}{I_{||}}=\ob,}
\end{equation}
where $\ob$ and $\oca$ are the  frequency of bounce oscillation along the magnetic field and the cyclotron frequency averaged over bounce oscillations.

For simplicity, let us assume a deeply trapped particle bouncing in the central part of a trap, where the magnetic field strength may be approximated with a parabolic function:
\begin{equation}\label{eqB}
\oc(z)=\occ\,\left(1+ z^2/(2l^2)\right).
\end{equation}
This is indeed a good approximation if fast electrons are generated in ECR discharge with the absorption zone located near the trap center. 
For the parabolic magnetic field, the Hamiltonian in the action-angle variables takes the following simple form:
\begin{equation*}
H_0=I_{||}\,\ob(I_\perp)+I_\perp\occ,
\end{equation*}
where 
\begin{equation*}\ob(I_\perp)=\sqrt{I_\perp\occ/(ml^2)}\end{equation*} 
is the non-relativistic bounce frequency. The averaged cyclotron frequency is 
\begin{equation*}\oca=\occ+(I_{||}/2I_\perp)\ob.
\end{equation*}
New angle coordinates are related to the old ones as
\begin{equation}\label{eq2}
\eqalign{\varphi=\xi_\perp-(I_{||}/4I_\perp)\sin 2\xi_{||},\cr
z=({2I_{||}/m\ob})^{1/2}\sin \xi_{||}.}
\end{equation}

Further we will use the relativistic Hamiltonian that may be obtained from the non-relativistic one as
\begin{eqnarray}\label{eqH0r}
\tilde{H}_0&=\sqrt{m^2c^4+2mc^2 H_0}=\nonumber\\ &\quad\quad\quad\quad=\sqrt{m^2c^4+2mc^2 (I_{||}\,\ob+I_\perp\occ)}\,.
\end{eqnarray}
Equations \eref{eq1} and \eref{eq2} remains valid for this Hamiltonian except corresponding relativistic frequencies are
\begin{equation*}\tilde\ob=\ob/\gamma,\quad\ocar=\oca/\gamma,
\end{equation*}
where $\gamma=\tilde H_0/mc^2$ is the standard relativistic factor.

\subsection{First-order perturbation}
Electron orbits in the presence of an unstable electromagnetic mode can be found from the perturbed Hamiltonian 
\begin{equation*}
H=\tilde H_0+H_1,\quad H_1=\frac{e}{mc\gamma}{\bi{A}\cdot\bi{p}},\end{equation*}
where $\bi{A}$ is the vector potential of the mode,  $\bi{p}$ is the kinematic momentum{, and the dot denotes the scalar product}. Note that only linear over $\bi{A}$ terms are accounted. Assuming that plasma mode is rotating with frequency $\o$ in the electron direction, the mode is standing over $z$, and the vector potential is transverse to the $z$-axis, one finds
\begin{eqnarray}
H_1=\frac{e}{mc\gamma}&\sqrt{2m(I_{||}\ob\sin^2\xi_{||}+I_\perp\occ)}\times\nonumber\\ &\quad\quad\times \,\mathrm{Re}[A_0(z)C(t)\exp(i\varphi-i\o t)].\label{eqH1}
\end{eqnarray}
Here $C(t)$ defines variation of the complex mode amplitude in time, $A_0(z)$ defines the fixed mode structure over $z$,  gyrophase $\varphi$ and coordinate $z$ are given by \eref{eq2}.

The perturbed Hamiltonian allows resonances of type \eref{res} at the fundamental cyclotron harmonic $s\!=\!1$. Higher cyclotron harmonics are not described because we omit finite {Larmor} radius effects in our two-dimensional Hamiltonian. To emphasize the resonances in explicit form, we expand the Hamiltonian into series over harmonics of the bounce oscillations:
\begin{equation}\label{eqHs}
	H=\tilde H_0+\mathrm{Re}\left[\sum_n V_nC(t)\exp(i\xi_\perp+in\xi_{||}-i\o t)\right]
\end{equation}
with
\begin{eqnarray*}\label{eqBn}
	V_n=\frac{e}{mc\gamma}\int_0^{2\pi} \sqrt{2m(I_{||}\ob\sin^2\xi_{||}+I_\perp\occ)}\times \\  \nonumber 
	\exp\left(-i\frac{I_{||}}{4I_\perp}\sin2\xi_{||}-in\xi_{||}\right)A_0\!\left(\sqrt{\frac{2I_{||}}{m\ob}}\sin \xi_{||}\right)\frac{\rmd\xi_{||}}{2\pi}.
\end{eqnarray*}

\subsection{Kinetic equation for a separate resonance}
Following Berk and Breizman\cite{brei97}, let us assume that resonances do not overlap. This means that spectral width $\Delta\omega$ of $C(t)$ and the effective collision frequency $\nu_\ind{eff}$ of resonant electrons (introduced later) are small compared with the bounce frequency, $\Delta\omega,\nu_\ind{eff}\ll\tilde\ob$. 

Then we may omit all terms in \eref{eqHs} except some $\bar{n}$ and change to new variables $\xi=\xi_\perp+\bar{n}\xi_{||}$ and $\xi_{||}$:
\begin{eqnarray*}
	H=&\sqrt{m^2c^4+2mc^2 \{(\bar{I}_{||}+\bar{n}I_\perp)\,\ob+I_\perp\occ\} }+\\ &\quad\quad\quad \quad\quad\quad 
	+\mathrm{Re}\left[ V_{\bar{n}}C(t)\exp(i\xi-i\o t)\right].
\end{eqnarray*}
Here $I_\perp$ and $\bar{I}_{||}={I}_{||}-\bar{n}I_\perp$ are the canonical momenta corresponding to coordinates $\xi$ and $\xi_{||}$. Obviously, $\bar{I}_{||}$ is conserved since $\xi_{||}$ do not enter the new Hamiltonian. Also, for resonant particles we may neglect variation of $I_\perp$ in the interaction term considering its value being close to the resonant value. 

Indeed, let us introduce the resonant frequency 
\begin{equation*}\Omega(I_\perp,\bar{I}_{||})\equiv\bar{n}\tilde\ob+\ocar\end{equation*} 
as a function of new actions:
\begin{equation}\label{eqO}
\Omega=\frac{(\bar{I}_{||}+3\bar{n}I_\perp)\,\ob+2I_\perp\occ}{2I_\perp\sqrt{1+2\{(\bar{I}_{||}+\bar{n}I_\perp)\,\ob+I_\perp\occ\}/mc^2}}.
\end{equation}
The equation of motion in $\xi$-direction, 
\begin{equation*}
\derv{^2\xi}{t^2}+\omega_\ind{w}^2\,\sin(\xi-\o t)=0,\quad
\omega_\ind{w}^2=\mathrm{Re}[V_{\bar{n}}C(t)]\pderv{\Omega}{I_\perp},
\end{equation*}
may be considered at fixed $I_\perp=\bar{I}_\perp$ corresponding to the resonance condition 
\begin{equation*}\Omega(\bar{I}_\perp,\bar{I}_{||})=\o.\end{equation*}
Quantity $\omega_\ind{w}$ defines the frequency of nonlinear oscillations of deeply trapped particles in wave $C$. Note, that following Berk and Breizman \cite{brei97} this quantity is frequently referred as ``bounce frequency'' $\ob$, but we use $\ob$ in a different sense.

With the same accuracy, we may formulate the kinetic equation for the distribution function $f(t,\xi,\Omega,\bar{I}_{||})$ of fast particles \cite{brei97}:
\begin{equation}\label{eqf}
\pderv{f}{t}+\Omega\pderv{f}{\xi}-\mathrm{Re}\left[iV_{\bar{n}}C(t)\,\rme^{i\xi-i\o t}\right]\pderv{\Omega}{I_\perp}\pderv{f}{\Omega}=\mathrm{St}f+Q.
\end{equation}
Here we have chosen $\Omega$ as a new independent variable to replace the action $I_\perp$. The forms of collision operator in the right-side of \eref{eqf} have been extensively discussed in \cite{brei97,berk99,lil09,lil10,les13}. Although the best choice in our particular case is still an open question, we chose the following model form
\begin{equation}\label{eqSt}
\mathrm{St}f+Q=\nu_\ind{diff}^3\pderv{^2}{\Omega^2}(f-F)+\nu_\ind{drag}^2\pderv{}{\Omega}(f-F),
\end{equation}
where $F$ is the equilibrium distribution, $\nu_\ind{dif}$ and $\nu_\ind{drag}$ are constants characterizing the momentum space diffusion and dynamical friction of fast electrons on the bulk plasma. Note, that \eref{eqSt} can be consistently derived from a Fokker--Planck collision operator with an appropriate orbit averaging \cite{Ref28}; then, up to numerical factors of the order of unity, 
\begin{equation}\label{eqSt1}\nu_\ind{diff}\approx(\o^2\nu_\ind{coll})^{1/3},\quad \nu_\ind{drag}\approx(\o\nu_\ind{coll})^{1/2}
\end{equation}
with $\nu_\ind{coll}$ being the pitch angle scattering rate. One finds that $\nu_\ind{drag}/\nu_\ind{diff}\approx(\nu_\ind{coll}/\o)^{1/6}\ll1$ in our conditions. However, solutions of the kinetic equation are rather sensitive to a small ratio $\nu_\ind{drag}/\nu_\ind{diff}$ \cite{lil10}, so we save the  dynamical friction for better fit to the experimental data. 

Solution of kinetic equation \eref{eqf} can naturally be found as a Fourier series over $\xi-\o t$,
\begin{equation*}
f-F=\sum_k f_k (t)\exp(ik\xi-ik\o t),
\end{equation*}
where $f_k(t)$ are slow functions compared to $2\pi/\o$.

\subsection{Equation for wave amplitude}
In the experiment, we see that instabilities develop near the discrete frequencies corresponding to low eigenmodes of an empty (with no plasma) cylindrical vacuum chamber \cite{viktorov_2016_EPL}. Besides, only modes with the transverse electric field near the axis are pronounced \cite{viktorov_2018_EPS}.	
Therefore, in the present communication we consider one of such modes with the transverse electric field
\begin{equation*}
\bi{E}=-\frac1c\pderv{\bi{A}}{t}\approx \frac{\o}{c}\,\mathrm{Re}\,[i \bi{A}_0(z,\bi{r}_\perp) C(t) \exp(-i\o t)],
\end{equation*}
rotating around the $z$-axis at frequency $\o$ in the same direction as electrons. Here $\bi{A}_0$ defines the vacuum mode structure, the mode is normalized such that 
\begin{equation}\label{eqAn}(\o/c)^2\int |\bi{A}_0|^2\rmd^3\bi{r}=1,\end{equation} 
and $C(t)$ is the slow amplitude of this mode.  
Following general procedure \cite{brei97}, an evolution of slow $C(t)$ may be determined from the Maxwell equations. In our case
\begin{equation}\label{eqC}
\derv{C}{t}=-\frac{2\pi i\o}{c}\,\rme^{i\o t}\int{\bi{A}_0^\dagger \cdot \bi{j}}\;\rmd^3\bi{r}-\gamma_\ind{d} C.
\end{equation}
The first term describes the Joule dissipation of total current $\bi{j}$ of resonant electrons; $^\dagger$ denotes a complex conjugate. The second term describes the external mechanism of wave dissipation with the damping rate $\gamma_\ind{d}$, e.g.\ due to diffraction and Ohmic losses from a cavity or by the background cold plasma.  

The first term in \eref{eqC} may be calculated explicitly having in mind that:
\begin{itemize}
	\item  the most significant contribution to the current of fast electrons is given from the resonant  Fourier harmonic $f_1$;
	\item the distribution function is normalized over a total number of particles in a flux tube, $\int f\,\rmd^4\Gamma=N$ with 
\begin{equation*}\rmd^4\Gamma=\rmd\xi_\perp\,\rmd I_\perp\,\rmd\xi_{||}\,\rmd{I_{||}}=\pderv{I_\perp}{\Omega}\,\rmd\xi\,\rmd \Omega\,\rmd\xi_{||}\,\rmd\bar{I_{||}};\end{equation*} 
	\item  a flux tube with resonant electrons is narrow compared to the diameter of the vacuum chamber, thus we may perform integration over the radial direction across the magnetic field just taking $\bi{j}\propto\delta(\bi{r}_\perp)$. 
\end{itemize}
With these assumptions,
\begin{equation*}
\rme^{i\o t}\int{\bi{A}_0^\dagger \cdot \bi{j}}\,\rmd^3\bi{r}\approx
\int{\frac{e}{m\gamma} f_1\,\rme^{-i\xi}\, \bi{A}_0^\dagger\cdot\bi{p}}\,\rmd^4\Gamma.
\end{equation*}
Next, the integrals over spatial coordinates $\xi$ and $\xi_{||}$ may be calculated. Finally,  equation \eref{eqC} for a slow complex wave amplitude takes the following form:
\begin{equation}\label{eqCf}
\derv{C}{t}=-{4\pi^2 i\o} \int V_{\bar{n}}^\dagger\,f_1\,\rmd I_\perp \rmd I_{||}-\gamma_\ind{d} C.
\end{equation}
As expected, an action of resonant particles on waves is defined by exactly the same function $V_{{n}}$ as one in the perturbed part of Hamiltonian \eref{eqHs}.

Note that $f_1$ describes both linear and nonlinear currents with respect to the wave amplitude $C$. For example, taking only linear and stationary perturbation $f_1^\ind{lin}$ from the kinetic equation \eref{eqf},
\begin{equation*}
i(\Omega-\o)f_1^\ind{lin}-iV_{\bar{n}}C\pderv{\Omega}{I_\perp}\pderv{F}{\Omega}=0,
\end{equation*}
and calculating a real part of the first term in \eref{eqCf}, one obtains the linear growth rate
\begin{eqnarray}
\gamma_\ind{L}={4\pi^2\o}\,&\mathrm{Im} \int \frac{1}{\Omega-\o} |V_{\bar{n}}|^2\pderv{\Omega}{I_\perp}\pderv{F}{\Omega}\,\rmd I_\perp \rmd I_{||}=\nonumber \\ \label{eqL}
&=4\pi^3{\o}\int \delta({\Omega-\o}){|V_{\bar{n}}|^2} \pderv{F}{\Omega}\,\rmd\Omega \,\rmd\bar{I}_{||}. 
\end{eqnarray}
This is the inverse Landau damping for the cyclotron waves in a collisionless plasma generalized over a case of inhomogeneous magnetic field.

``Holes and clumps'' structures are determined by a nonlinear contribution to the current what requires self-consistent calculation of slow evolution of at least $f_0$, $f_1$ and $f_3$ taking into account the following ordering $F\gg f_1 \gg f_0\sim f_2\gg f_3...$. We find such solutions numerically using the BOT code by M. K. Lilley \cite{BOT}.  

\subsection{Limit of constant magnetic field}
To check a consistency of our theory we consider $l=\infty$. In this case, the magnetic field is homogeneous and there is no bounce oscillations: $\oc=\occ$, $\ob=0$ and $I_{||}=0$. Unperturbed particle orbits correspond to a constant parallel momentum. The coupling parameter is  
\begin{equation*}
V_n=\frac{e}{mc\gamma} \sqrt{2m I_\perp\occ}A_0\int_0^{2\pi}\!\!\!\rme^{-in\xi_{||}}\frac{\rmd\xi_{||}}{2\pi}=\frac{e}{c} A_0 v_\perp \delta_{0n}, \quad  \end{equation*}
The resonant frequency is  equal to the relativistic gyrofrequency, $\Omega=\oc/\gamma$ with $\gamma=\sqrt{1+2I_\perp\oc/mc^2}$. Then, the nonlinear trapping frequency is
\begin{equation*}
\omega_\ind{w}^2=\mathrm{Re}\,C V_0\derv{\Omega}{I_\perp}= \frac{v_\perp}c \frac{E}{B}\frac{\oc^2}{\gamma^2}, \end{equation*}
here $E$ is a wave electric field amplitude, $B$ is the external magnetic field strength. This is exactly the trapping (bounce) frequency for a resonant electron in the slow extraordinary wave in a rarefied plasma \cite{trakh_84}.

Within the same assumptions one obtains the linear growth rate \eref{eqL} as
\begin{equation}\label{eqL0}
\gamma_\ind{L}=\frac{8\pi^3 e^2c^2}{\oc^2V}\left(\gamma^2 I_\perp \pderv{F}{I_\perp}\right)_{\,\Omega=\o}, 
\end{equation}
where $V$ is the effective volume used in norm \eref{eqAn},  $(\Omega/c)^2A_0^2V=1$. On the other hand, a theory of homogeneous plasma results in the following standard expression for the cyclotron (Landau) damping/growth rate of the extraordinary plasma mode propagating transverse to the magnetic field at the fundamental  harmonic, 
\begin{eqnarray*}
\gamma_\ind{L}^0=\pi^2\frac {\omega_\ind{p}^2}{\oc}\int \delta(\oc/\gamma &-\o)\,\frac{p_\perp^2}{\gamma} \pderv{f}{p_\perp}\,\rmd p_\perp=\\
&={\pi^2} \frac{\omega_\ind{p}^2}{\oc}m^2c^2\left(\gamma^2 p_\perp \pderv{f}{p_\perp}\right)_{\,\Omega=\o} 
\end{eqnarray*}
with $\omega_\ind{p}^2=4\pi e^2 N/mV$ being the Langmuir frequency of fast electrons (see, e.g., \cite{Melrose1980}). One finds that both equations are equivalent, $\gamma_\ind{L}\equiv\gamma_\ind{L}^0$, with taking into account different norms of distribution functions
\begin{equation*}
\int F\,2\pi \rmd I_\perp=N,\quad \int f\,2\pi p_\perp \rmd p_\perp=1.\end{equation*}

\section{Interpretation of the experiment}

\subsection{Experimental conditions and basic parameters}
For reader's convenience, we remind the experimental conditions in which the frequency chirping has been systematically observed \cite{viktorov_2016_EPL, viktorov_2018_EPS, shalash_2017_pop}. Most relevant parameters to our study are summarized in \tref{tab1}.

\begin{table}[b]
\caption{ \label{tab1} Typical parameters for the stationary ECR discharge and the plasma decay stage after  ECR heating switch-off at which the frequency chirps are observed at SMIS-37 setup.}
\begin{indented}\item[]
\begin{tabular}{@{}lll}
	\br
	& \textbf{ECRH}  & \textbf{Decay}   \\
	\mr
	Background plasma density \rem{$N_{\ind{c}}$}&  $\sim 10^{13}$ cm$^{-3}$  & $\lesssim 10^{11}$ cm$^{-3}$\\
	Fast el.\ density (1--30 keV)\rem{$N_{\mathrm{h}}$} & $10^{11}$ cm$^{-3}$ & $10^{11}$ cm$^{-3}$\\
	Fast el.\ density ($>$100 keV)\rem{$N_{\mathrm{h}}$} & $10^{9}$ cm$^{-3}$ & $10^{9}$ cm$^{-3}$\\
	Bulk electron temperature \rem{$T_\ind{c}$}&  up to 300 eV  & $\sim 1$ eV \\ 
	Fast electron energy (ave.) \rem{$\varepsilon_{\mathrm{h}}$} & $ 10$ keV &  $10$ keV \\ 
	Fast electron energy (max.) \rem{$\varepsilon_{\mathrm{h}}^*$} & $350$ keV &  $300$ keV \\ 
	\mr
	Heating frequency\ $\omega_\ind{ECH}/2\pi$ & 37.5 GHz& --\\
	Min.\ cyclotron freq.\  $\occ/2\pi$ & 10 GHz & 8 GHz\\ 
	Electron collision rate $\nu_\ind{coll}$ & &\\
	\qquad  background plasma & $5\cdot10^5$ s$^{-1}$ & $10^7$ s$^{-1}$ \\
	\qquad  300 keV electrons & 10 s$^{-1}$ & 0.1 s$^{-1}$ \\
	\br
\end{tabular}
\end{indented}\end{table}

The experiments were conducted in the plasma of ECR discharge sustained by gyrotron radiation launched along the magnetic field in the simple axially symmetric open magnetic trap. Plasma was created and supported under ECR conditions at the fundamental  harmonic (frequency 37.5\,GHz, power up to 80\,kW, pulse duration  1\,ms). Two resonance surfaces were located symmetrically quite far, by 7.5 cm, from the trap center towards the magnetic mirrors. 

For the needs of the present work, the on-axis magnetic field may be approximated as
\begin{equation*}
\oc(z)={\occ}\,{\left(\frac{1+R}2+ \frac{1-R}{2}\cos\frac{2\pi z}L\right)},
\end{equation*}
with the trap length $L=32$\,cm and mirror ratio $R=5$. This corresponds to the magnetic field scale $l=3.6$\,cm in \eref{eqB} calculated as $l=L/(\pi\sqrt{2(R-1)})$.

Highly transient spectra with many repeated frequency sweeps were detected with the use of a broadband 2--20\,GHz horn antenna and a high-performance oscilloscope. The frequency chirps  were observed only at the late stage of plasma decay with a pronounced delay after the ECR heating switch-off (0.1--1\,ms) and only when the ambient magnetic field was decreasing in time. The microwave emission was observed only at frequencies below the electron cyclotron frequency at the trap center in a few  narrow frequency bands, 4.9--5.3\,GHz, 6.2--6.5\,GHz, 6.9--7.1\,GHz, 7.2--7.3\,GHz and 7.5--7.6\,GHz.
{With a good accuracy, these frequency bands are linked to the eigenmodes of the vacuum chamber; therefore, the bands do not depend on gas discharge conditions, such as gas pressure, ECRH power and even the magnetic field strength. However, when the magnetic field  decreases in time such that the fundamental cyclotron harmonic $\occ/2\pi$ becomes lower than the corresponding frequency band, the microwave emission is not detected anymore in  this particular band but still may  be detected at lower frequency bands.}
Within each frequency band, the dynamic spectrum represents a set of chirped bursts with the duration of about 10\,$\mu$ and both increasing and decreasing frequencies in the range of 100--400\,MHz depending on the experimental conditions. 
A typical example of such an event in a nitrogen discharge is shown in \fref{fig1}.

As mentioned above, a vacuum chamber forms a closed high-$Q$ cavity for a potentially unstable plasma wave. A basic electrodynamic analysis suggests that the unstable mode is a standing wave with a circularly polarized transverse electric field near the trap axis. Namely, the frequency bands mentioned above may be attributed to following modes of a cylindrical resonator: TE$_{211}$, TE$_{311}$, TE$_{312}$, TM$_{310}$ and TE$_{411}$ \cite{viktorov_2018_EPS}.	 However, for the present study, such details are even excessive. Indeed, in our experimental conditions, due to a localized heating, most of fast electrons are bouncing near the trap center far from the end-fronts of the vacuum chamber. As a first approximation we may neglect longitudinal structure of the electromagnetic mode simply assuming $A_0=\mathrm{const}$ when calculating the coupling coefficient $V_n$. So, the particular mode type enters our calculations only through the norm \eref{eqAn} of $\bi{A}_0(\bi{r})$.

\subsection{Model distribution function}

Measurements of an energy distribution function of fast electrons leaving our trap through the mirrors were reported in \cite{Izotov_2012_RSI}. 
The electron energy distribution function was found to decrease slowly with energy up to approximately 500\,keV and drops abruptly further thus being very far from the Maxwellian one. Basing on these results and noting that direct measurements of the full distribution function inside a high-power ECR discharge are not possible with the present diagnostics, we propose the following physical model. 

During the ECR heating stage, the interaction between resonant electrons and a high-power external monochromatic wave is well described within the quasilinear theory. In the process of heating, the resonant electrons redistribute along the quasilinear diffusion curve which in our notations is $\bar{I}_{||} =\mathrm{const}$. An almost flat energy spectrum of escaping electrons indicates that the wave intensity in our experiment is so strong that the quasilinear diffusion results in almost a plateau along the curves $\bar{I}_{||} =\mathrm{const}$ lasting from low energies of few keV up to the maximum energy $\epsilon^*=mc^2(\gamma^*-1)$.
Thus, the initial non-perturbed distribution function is effectively one-dimensional: $F=\mathcal{F}(\bar{I}_{||})$ for $\gamma <\gamma^*$ and zero for higher energies. Function  $\mathcal{F}$ may be found by matching to the Maxwellian distribution function at low energies.

\def\C{\mathcal{K}}
\def\a{\alpha}
\def\Te{T_\ind{e}}
Now let us stress one technical detail. In our study we must distinguish the magnetic field during the ECR discharge and the instability stage. Thereafter we assume that $\occ$ and $\ob$ characterize the magnetic field strength and the bounce oscillation rate \emph{at the instability start} and  $\occ^*=\a\occ$ and $\ob^*=\sqrt{\a}\ob$ characterize the same quantities \emph{during the EC heating}. In our experiment, the decompression factor $\a$ varies from 1.25 to 2. 

Having in mind that formation of the electron distribution function occurs during the heating stage, we find that conservation of $\bar{I}_{||}$ is equivalent to conservation of 
\begin{eqnarray}\label{eqK}
\C(I_\perp,I_{||})=mc^2(\gamma-1)-I_\perp\omega_\ind{ECH}= \\ \nonumber =\sqrt{m^2c^4+2mc^2(I_{||}\ob^*+ I_\perp\occ^*)}-I_\perp\omega_\ind{ECH}-mc^2,
\end{eqnarray}
where $\omega_\ind{ECH}$ is the heating frequency. This definition of the quasilinear diffusion line may be reformulated in a more conventional form (see, e.g.,  \cite{shalNF} and references therein)
\begin{equation*}
\C=mc^2(\gamma-1)- p'^2_\perp/2m\approx p'^2_{||}/2m,\end{equation*}
where $p'_\perp$ and $p'_{||}$ are the transverse and longitudinal to the magnetic field components of the electron momentum calculated at the position of ``cold'' EC resonance $\oc(z)=\omega_\ind{ECH}$; the approximate equality corresponds to the well-known non-relativistic limit (low energy electrons gain presumably transverse momentum during the EC heating). Assuming that accelerated electrons originate from a relatively cold and dense background  plasma, we may consider a Maxwellian distribution over longitudinal momentum for the seed electrons, $\mathcal{F}\propto\exp(- p'^2_{||}/2m\Te)$, with some effective temperature $\Te$. This condition, formulated at low energies, defines the spread of $\C$ in the whole relativistic domain. Summarizing, we define the following model distribution function
\begin{equation}\label{eqfm}
F=F_0\exp\left(-\frac{\C(I_\perp,I_{||})}{\Te}\right)\cases{1 &$\!\!\!\!\!\!\!\!\!\!$for $\gamma<\gamma^*$\\ 0&$\!\!\!\!\!\!\!\!\!\!$for $\gamma>\gamma^*$},\end{equation}
where $F_0$ is defined through the norm,  
\begin{eqnarray}
N&=\int F\,\rmd^4\Gamma=(2\pi)^2\int F\, \derv{I_{||}}{\C}\rmd\C\rmd I_\perp\approx \label{eqnorm} \\ 
&\approx (2\pi)^2F_0\int_0^{\infty}\rme^{-\C/\Te}\rmd \C \int_0^{I_\perp^*}\left(\derv{I_{||}}{\C}\right)_{\C=0}\rmd I_\perp= \nonumber\\
&=8\pi^2F_0\frac{\Te}{mc^2}\frac{(mc^2+\case13\epsilon^*)\,\epsilon^*}{\ob^{**}\,\omega_\ind{ECH}}\nonumber,\end{eqnarray} 
$I_\perp^*=\epsilon^*/\omega_\ind{ECH}$ and $\ob^{**}=\sqrt{\a}\ob(I_\perp^*)$. Here we use the following approximation. The effective temperature of the bulk electrons,  $\Te\sim100-300$ eV, is much less than the energy of the resonant electrons involved in the instability. Therefore, in most cases, we may neglect the spread in $\C$, formally this corresponds to change $\exp(-\C/\Te)/\Te\to\delta({\C})$ at $\Te\to0$. We do this when define $\dervi{I_{||}}{\C}$ -- evidently this quantity depends only on $I_\perp$ if we imply $\C=0$, and then all integrals may be calculated. The same trick will be used next to determine the bounce resonances, however to avoid singularities we must retain finite $\Te$ in all kinetic calculations (both linear and nonlinear). 

Our model allows determination of the upper limit $\epsilon^*$ for the electron energy. To do this, let us remind that all particles adiabatically confined in a magnetic mirror must satisfy the following condition:
\begin{equation}\label{eqlc}
\frac{p_{\perp0}^2}{p^2}\equiv\frac{I_\perp \occ^*}{I_\perp \occ^*+\ob^*I_{||}}\ge\frac{B_{\min}}{B_{\max}}\equiv\frac1R,\end{equation}
where index ``0'' refers to the trap center.  Assuming  $\Te\to0$ and solving inequality \eref{eqlc} along the quasilinear diffusion curve $\C(I_\perp,I_{||})=0$, we find that it is equivalent to
\begin{equation*}
mc^2(\gamma-1)\le\epsilon^*\equiv 2mc^2\,(R \occ^*/\omega_\ind{ECH}-1).\end{equation*}
Thus, once a particle following the diffusion curve acquire an energy higher than $\epsilon^*$, it is lost through the loss-cone. For our particular experimental conditions, we may estimate the maximum electron energy gained during a strongly off-center ECR heating as $\epsilon^*\approx 340$ keV, which is in reasonable agreement with the measurements.

On a time interval between the ECH and chirp instability, the Coulomb collisions of fast electrons are negligibly weak. Therefore, the action variables are adiabatic invariants during slow (compared to the bounce and cyclotron motions) plasma decay while the magnetic field and the bulk plasma vary essentially. This means that the distribution function \eref{eqfm} remains the same during of the instability stage or, at least, at the beginning of this stage. In particular, during the decay stage the maximum electron energy diminishes as approximately $\epsilon\propto\sqrt{B}$.

\subsection{Resonances}

Our model distribution function may be used to determine possible resonances between electrons and an unstable plasma mode. In the limit of $\Te\to0$, such resonances are determined by the following conditions:
\begin{equation*}
\Omega(I_\perp,I_{||})=\o,\quad \C(I_\perp,I_{||})=0, \quad I_\perp<I_\perp^*.\end{equation*}
Simultaneous solution of the first two equations gives actions $I_\perp$ and  $I_{||}$ as a function of the resonance order ${n}$, the last inequality gives the upper limit for ${n}$. Correspondingly, one may calculate all functions of the actions like particle energy, bounce frequency etc. 

\Tref{tab2} presents an example of such calculations for the experimental spectrum shown in \fref{fig1}. Note that in a symmetric trap only even $n$ may be excited because the coupling $V_n$ is exact zero for all odd $n$. 
One can see that:
\begin{itemize}
	\item the bounce frequencies $\tilde\ob$ correspond to the observed frequency sweep in a single burst;
	\item the neighboring bounce frequencies do not overlap,  $\Delta\tilde\ob>\nu_\ind{coll},\gamma_\ind{L}$;
	\item different  ${n}$ correspond to groups of resonant electrons with different energies;
  \item the resonance is possible only for negative $n$, thus maximum electron energy $\epsilon^*$ defines the \emph{minimal} allowed resonance order; in our case ${|n|}\ge 34$;
	\item coordinate $z_\ind{loc}$, defined as a solution of $\oc(z_\ind{loc})/\gamma=\o$, characterizes a spatial position of the emitting zone -- the wave-particle interaction is indeed located near the trap center.

\end{itemize}
Next we will describe the linear growth rate, also presented in  \tref{tab2}.

\begin{table}[tb]
\caption{ \label{tab2} Resonant values of the particle energy $\epsilon=mc^2(\gamma-1) $, the linear growth rate $\gamma_\ind{L}$, the bounce frequency $\tilde\ob$, the local cyclotron frequency $\oc(z_\ind{loc})=\gamma\o$ and its coordinate for successive  resonance orders ${n}$. Parameters correspond to experimental data shown in \fref{fig1}: $\o/2\pi=6.25$\,GHz, $\occ/2\pi=8$\,GHz at the start of chirp activity and $\occ^*/2\pi=10$\,GHz during the ECR heating at $\omega_\ind{ECH}/2\pi=37.5$\,GHz. Maximal electron energy is $\epsilon^*=340$\,keV during the heating; due to magnetic decompression it reduces down to 305\,keV when the instability develops. The coupling parameter varies from $|V_n|/eA_0=0.07$ at $n=-30$ to 0.04 at $n=-42$.  }
\begin{indented}\item[]
\begin{tabular}{@{}llllllll}
	\br
	${n}$& --30 &--32 &\textbf{--34}  & --36 & --38 & --40 & --42 \\
	\mr
$\epsilon$, keV & 380 & 320 & \textbf{270} & 235 & 205 & 180 & 160\\
$\gamma_\ind{L}$, $\times10^{7}$s$^{-1}$ & 12 &7.2 & \textbf{3.6} & 1.5 & 0.3 & --0.3 &--0.5\\
$\tilde\ob/2\pi$, GHz& 0.54 & 0.50 &\textbf{0.46} & 0.43 & 0.40 & 0.38 & 0.36 \\
$\oc/2\pi$, GHz& 10.8 & 10.1 & \textbf{9.5} & 9.1 & 8.7 & 8.4 & 8.2\\
$z_\ind{loc}$, cm & 2.8 & 2.4 & \textbf{2.0} & 1.8 & 1.5 & 1.0 & 0.8\\
	\br
\end{tabular}
\end{indented}\end{table}

\subsection{Linear growth rates}

The model distribution function \eref{eqfm} represents a plateau along quasilinear diffusion lines, $(\pdervi{F}{I_\perp})_{\bar{I}_{||}}=0$, what guarantees absence of the linear interaction, $\gamma_\ind{L}=0$, at the heating frequency $\omega_\ind{ECH}$ and the magnetic field $\occ^*$. During the instability phase both the magnetic field and radiation frequency (of an unstable mode) are different, so the diffusion lines are different too and $(\pdervi{F}{I_\perp})_{\bar{I}_{||}}\ne0$.  In other words, $\C$ as defined by \eref{eqK} is a local function of ${\bar{I}_{||}}$ only for the heating radiation during the heating phase. When we consider the instability phase, $\C$ and ${\bar{I}_{||}}$ may be treated as independent variables. This observation allows characterizing the inverse Landau damping in an explicit way.

Indeed, after substitution of \eref{eqfm} into \eref{eqL},  change of variable $I_\perp$ to $\C$, and integration over ${\bar{I}_{||}}$, we find
\begin{eqnarray*}
\gamma_\ind{L}&=4\pi^3\sigma{\o}\int \delta({\Omega-\o}){|V_{\bar{n}}|^2} \pderv{F}{\C}\,\rmd\C \,\rmd\bar{I}_{||}= \\
&=4\pi^3\sigma{\o}\int\! { |V_{{n}}|^2\left|\pderv{\Omega}{\bar{I}_{||}}\right|^{-1} } \pderv{F}{\C} \,\rmd{\C}=\\
&=-4\pi^3\sigma{\o}\int\! \pderv{}{\C}\left\{ {|V_{{n}}|^2 \left|\pderv{\Omega}{\bar{I}_{||}}\right|^{-1} } \right\}F \,\rmd{\C}\end{eqnarray*} 
$\sigma= \mathrm{sign}\left(\pdervi{\C}{I_\perp}\right)_{\bar{I}_{||}}$ and  $(\pdervi{\Omega}{\bar{I}_{||}})_{\C}$ are calculated with a help of \eref{eqO} and \eref{eqK}. For small $\Te$, one may consider $F$ as $\delta$-function, then the last expression may be calculated  as
\begin{equation*}
\gamma_\ind{L}\approx-4\pi^3\sigma{\o}\pderv{}{\C}\left\{{|V_{{n}}|^2 \left|\pderv{\Omega}{\bar{I}_{||}}\right|^{-1} } \right\}_{\!\C=0}\int\! F \,\rmd{\C}.\end{equation*} 
Noting that $\int\! F \rmd{\C}=F_0 \Te$ and using explicit expression \eref{eqnorm} for the norm,   we finally find
\begin{equation}\label{eqL3}
\gamma_\ind{L}\approx -\frac{\pi \sigma}{2} \frac{Nmc^2\o\ob^{**}\,\omega_\ind{ECH}}{(mc^2+\case13\epsilon^*)\,\epsilon^*} |V_{{n}}|^2 \left\{ { \pderv{}{\C}\left|\pderv{\Omega}{\bar{I}_{||}}\right|^{-1} } \right\}_{\!\C=0}.
\end{equation} 
In our model, the linear instability growth rate is determined by a topology of the diffusion and resonance curves through $\sigma$ and $\pdervi{\Omega}{\bar{I}_{||}}$, but it is independent of the spread of the distribution function $\Te$ (this is only a technical parameter introduced to avoid singularities).

We find that in our experimental conditions $\sigma<0$. Indeed, after some algebra this condition may be formulated as 
$\sqrt{\a}\,\o+(\a-\sqrt{\a})\,\occ/\gamma<\omega_\ind{ECH}$, 
what is always true for large enough $\omega_\ind{ECH}/\o$ and $\a\sim1$. 
Therefore, the instability corresponds to the positive sign of the term in curly brackets in \eref{eqL3}.

An example of the linear instability growth rate $\gamma_\ind{L}$ for different resonances is presented in \tref{tab2}.
We find that only harmonics with $20\le|n|\le38$ are unstable. On the other hand, harmonics with {$|n|<34$} correspond to resonant energies above our upper limit for the electron energy, $\epsilon^*$. Therefore, only few harmonics, $n=-34,-36,-38$, can contribute to the ``holes and clumps'' dynamics.  
Next we will analyze in more details the lowest possible resonance $n=-34$ corresponding to the maximal linear growth rate.

Our results agree with the previous analysis of the experimental data resulted in the range of instability linear growth rates $(1-6)\times 10^7$\,s$^{-1}$and the energy of emitting electrons of 170\,keV  \cite{viktorov_2016_EPL}. Higher electron energies indicated in \tref{tab2} are related to the variation of the magnetic field being taken into account: in the new model electrons are not emitting exactly in minimum-$B$ thus they must have an additional relativistic downshift of the cyclotron frequency.

\subsection{Numerical simulation of frequency chirps}

One of results of the present theory is that master equations \eref{eqf}, \eref{eqCf} for the cyclotron interaction are exactly the same as those considered earlier for the case of interaction of particles with electrostatic plasma waves. Thus, we may use one of many numerical tools developed for 1D electrostatic simulations. At this, all details related to a particular electromagnetic mode structure and mechanism of particle-wave interaction are taken into account in norming factors and dimensionless parameters of calculations. 

In this paper, we find a numerical solution of \eref{eqf},\eref{eqCf} with open source code BOT originally developed to study bump-on-tail instabilities \cite{BOT, lil10}.  
There are four main physical input parameters of the code: 
\begin{itemize}
	\item the initial wave amplitude{, $E\propto\omega_\ind{w}^2$,  characterized by the dimensionless parameter} $\omega_\ind{w}/\gamma_\ind{L}$; 
	\item the wave damping rate {measured} as $\gamma_\ind{d}/\gamma_\ind{L}$;
	\item the collisional diffusion rate as $\nu_\ind{diff}/(\gamma_\ind{L}-\gamma_\ind{d})$; 
	\item the collisional drag rate as $\nu_\ind{drag}/(\gamma_\ind{L}-\gamma_\ind{d})$.
\end{itemize}
We do not use the code option for the Krook collision operator. All output  are given in units of the linear growth rate $\gamma_\ind{L}$, thus we have an additional parameter when fitting the result to the experimental data in dimensional units.

\begin{figure}[tb]
\centering
\includegraphics[width=83mm]{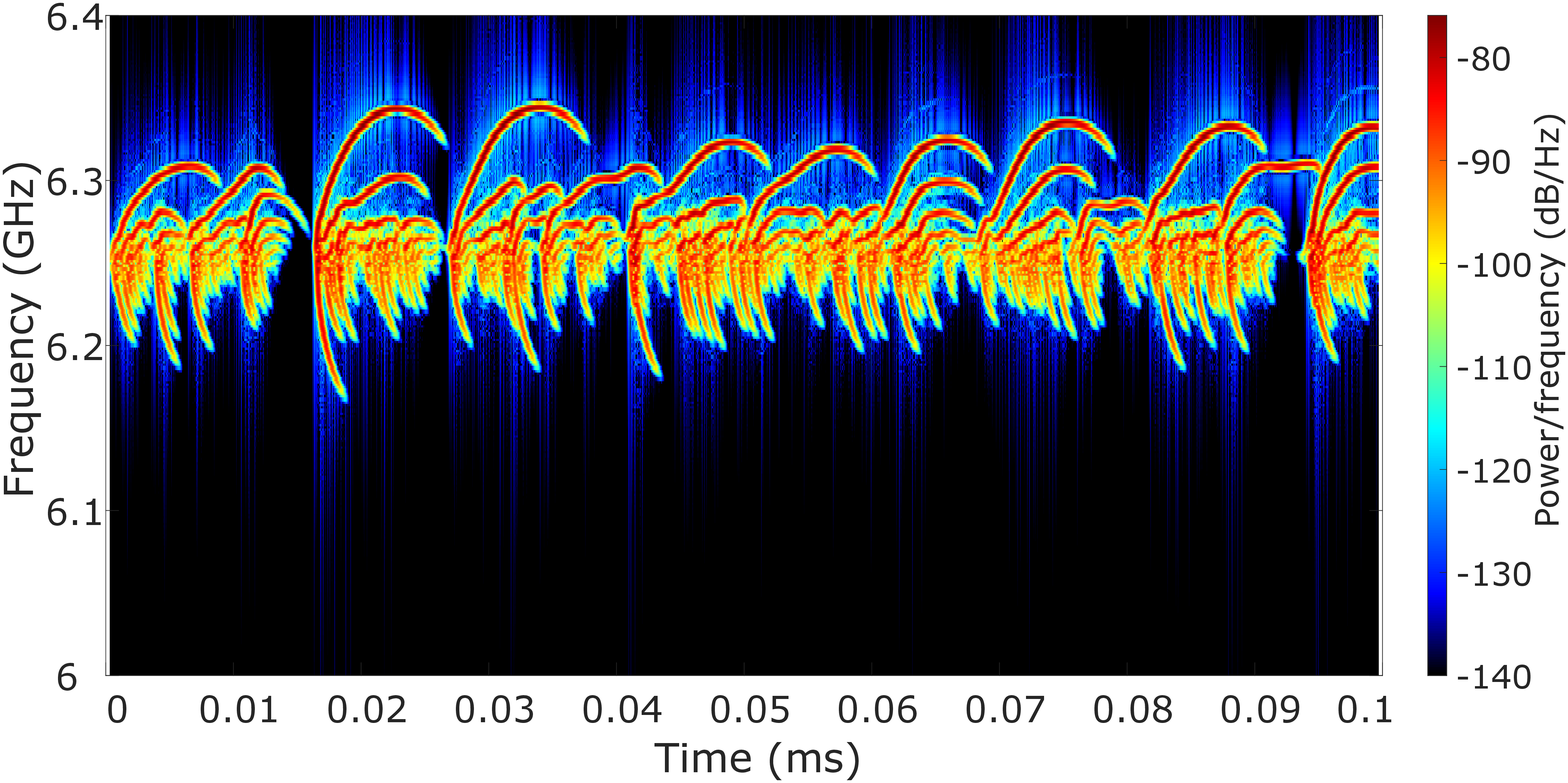}
\includegraphics[width=83mm]{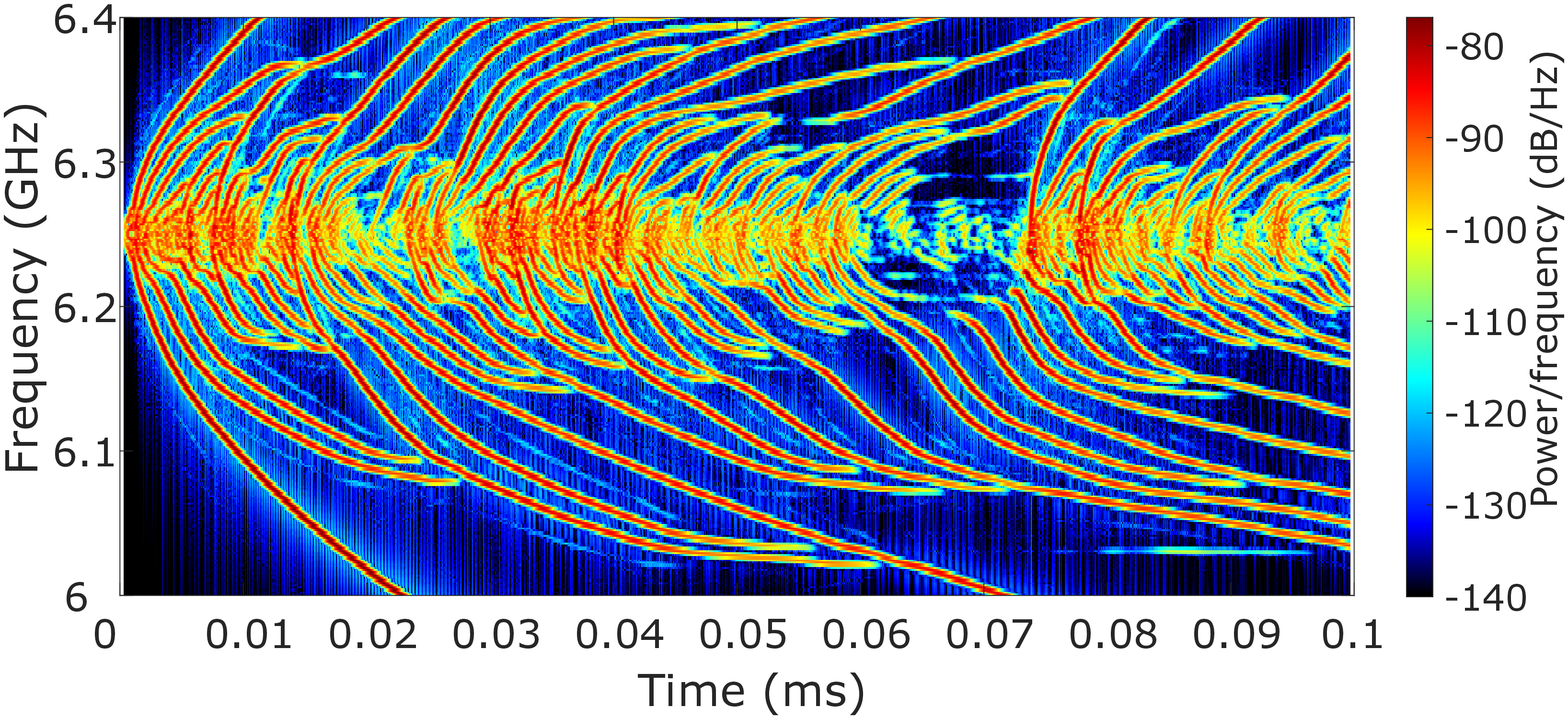}
\caption{Dynamic spectra of stimulated electron cyclotron emission calculated with BOT code. Top panel: result of full optimization, corresponding parameters  are $\omega_\ind{w}/\gamma_\ind{L}=0.1$, $\gamma_\ind{d}/\gamma_\ind{L}=0.92$, $\nu_\ind{diff}/(\gamma_\ind{L}-\gamma_\ind{d})=1.4$, $\nu_\ind{drag}/(\gamma_\ind{L}-\gamma_\ind{d})=1.6$. Bottom panel: result of partial optimization for the fixed collisional drag force, $\nu_\ind{diff}/(\gamma_\ind{L}-\gamma_\ind{d})=0.2$, $\nu_\ind{drag}/(\gamma_\ind{L}-\gamma_\ind{d})=0.02$, other parameters are the same as for the top plot.    We use 10 harmonics in terms of BOT code manual \cite{BOT}. Plasma conditions correspond to \fref{fig1}.  }
\label{fig2}
\end{figure}

\Fref{fig2} shows two  examples of simulation of the particular series of observed frequency chirps presented in \fref{fig1}. 
To plot this figure we use the following procedure.

The code describes a singular resonance, so we select $n=-34$ resonance considered in the previous section. Then we have an initial guess for the linear wave growth rate $\gamma_\ind{L}$, see \tref{tab2}, and for the collisional rates, $\nu_\ind{diff}=(\o^2\nu_\ind{coll})^{1/3}$ and $\nu_\ind{drag}=(\o\nu_\ind{coll})^{1/2}$,  defined by the energy of resonant electrons $\epsilon$. At this stage we assume zero collisional drag. 

\begin{table}[tb]
\caption{ \label{tab3} Physical parameters for BOT simulations.  }
\begin{indented}\item[]
\begin{tabular}{@{}lll}
	\br
	Parameter& Initial guess & After adjustment \\
	\mr
$\gamma_\ind{L}$ & $3.6\times10^{7}$ s$^{-1}$ & $7.0\times10^{7}$ s$^{-1}$ \\
$\gamma_\ind{d}$ & $3.45\times10^{7}$ s$^{-1}$ & $6.44\times10^{7}$ s$^{-1}$ \\
$\nu_\ind{diff}$ & $5.3\times10^{6}$ s$^{-1}$ & $7.8\times10^{6}$ s$^{-1}$ \\
$\nu_\ind{drag}$ & $6.3\times10^{4}$ s$^{-1}$ & $9.0\times10^{6}$ s$^{-1}$ \\
	\br
\end{tabular}
\end{indented}\end{table}

The initial wave amplitude is defined by the ``thermal'' fluctuations in a hot plasma \cite{Krall}; its level may be estimated as 
$|E|^2/8\pi\sim\epsilon\o \Delta\omega/\pi c^3$
with  $\o\approx 6.2$ GHz, $\Delta\omega/2\pi\approx 100$ MHz, and $\epsilon\approx 270$ keV. With these data we get $\omega_\ind{w}/\gamma_\ind{L}\approx 0.1$. 

The most uncertain parameter is the damping rate $\gamma_\ind{d}$. Since the observed chirp series looks quite stationary during times comparable with the plasma decay rate, we conclude that $\gamma_\ind{d}$ is more likely related to losses from a vacuum chamber than to wave dissipation in the background plasma\footnote{Note that in \cite{viktorov_2016_EPL}, following our previous experience for other types of instability \cite{shalash_2006,shalash_jetpl,shalash_ppcf}, we assume that the instability threshold is determined by dissipation in the background plasma. Collecting more experimental data force us to change this  assumption, for a particular case of chirp emission,  to the opposite one.}. Due to many diagnostic ports, it is hard to estimate \textit{a-priory} a real $Q$-factor of our  vacuum chamber. On the other hand, we know that   Berk--Breizman model is very sensitive to the ratio $\gamma_\ind{d}/\gamma_\ind{L}$, thus it may be recovered by fitting of the code output to the experimental data.  As a first guess, we assume that the instability starts close to the threshold, $\gamma_\ind{d}/\gamma_\ind{L}=0.96$. Here we remind that complex transient dynamics is possible only if $\nu_\ind{diff}/(\gamma_\ind{L}-\gamma_\ind{d})<\hat\nu_\ind{cr}=4.38$ \cite{berk96}; in our case this is equivalent to  $\gamma_\ind{d}/\gamma_\ind{L}<0.967$.

After initial guess is done, we run BOT code several times and adjust input parameters to achieve a better fit of the calculated and experimental dynamical spectra. 
\Fref{fig2}(top) shows the result of such optimization for a particular case shown in  \fref{fig1}; corresponding parameters in physical units are listed in \tref{tab3}. 
Here we match rather complex 2D images, but the procedure is fairly straightforward: roughly, scaling of the time axis scale refines $\gamma_\ind{L}$, scaling of the frequency axes refines $\gamma_\ind{d}$, and matching the whisker pattern refines $\nu_\ind{diff}$ and $\nu_\ind{drag}$. During fitting only one parameter, $\nu_\ind{drag}$, changes essentially compared to its theoretical estimate. For reference, in  \fref{fig2}(bottom)  we also present the result of a simulation in which we optimize all parameters except $\nu_\ind{drag}$ (initial guess for the collisional drag force is fixed).

\begin{figure}[tb]
\centering
\includegraphics[width=83mm]{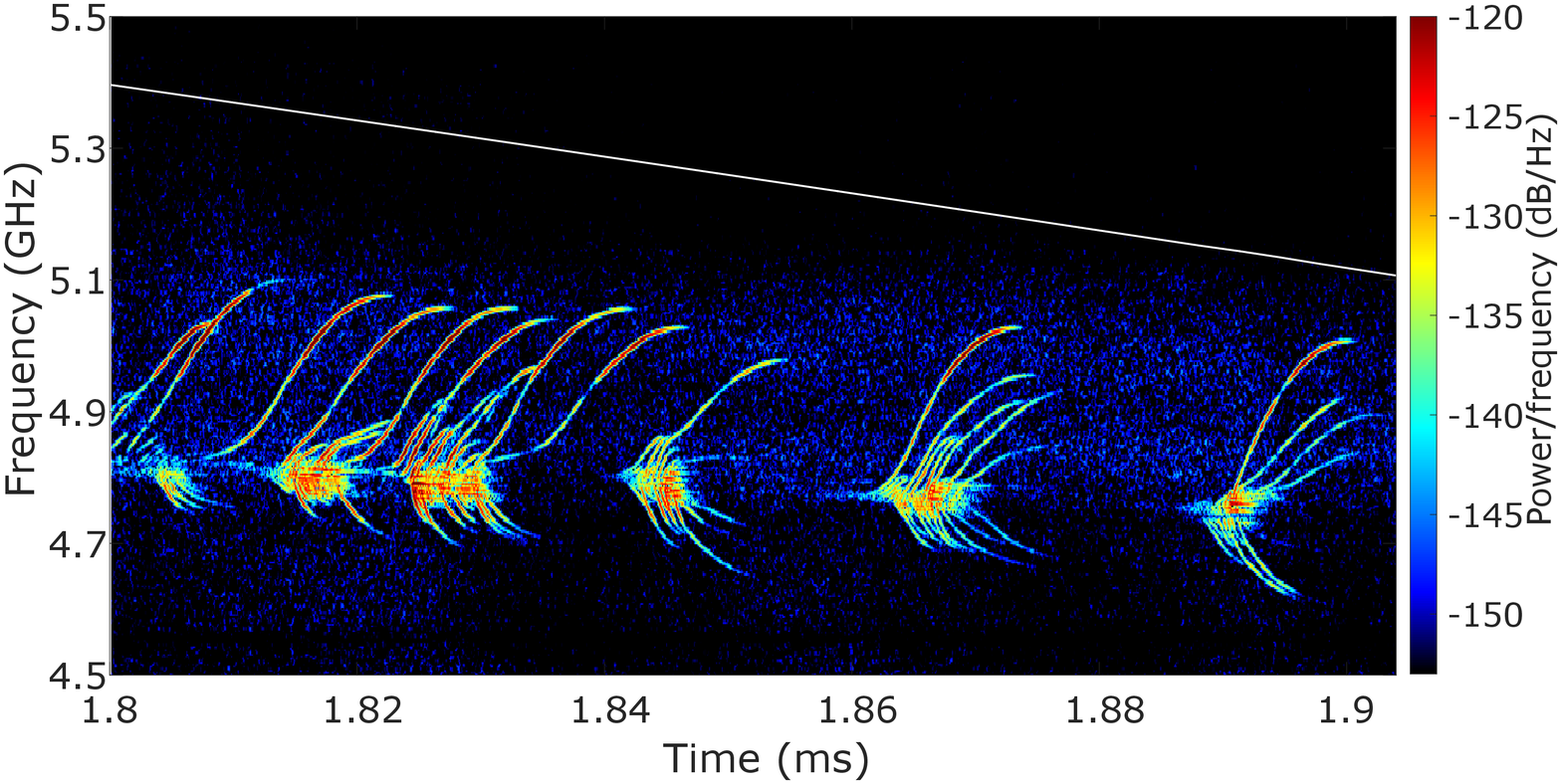}
\includegraphics[width=83mm]{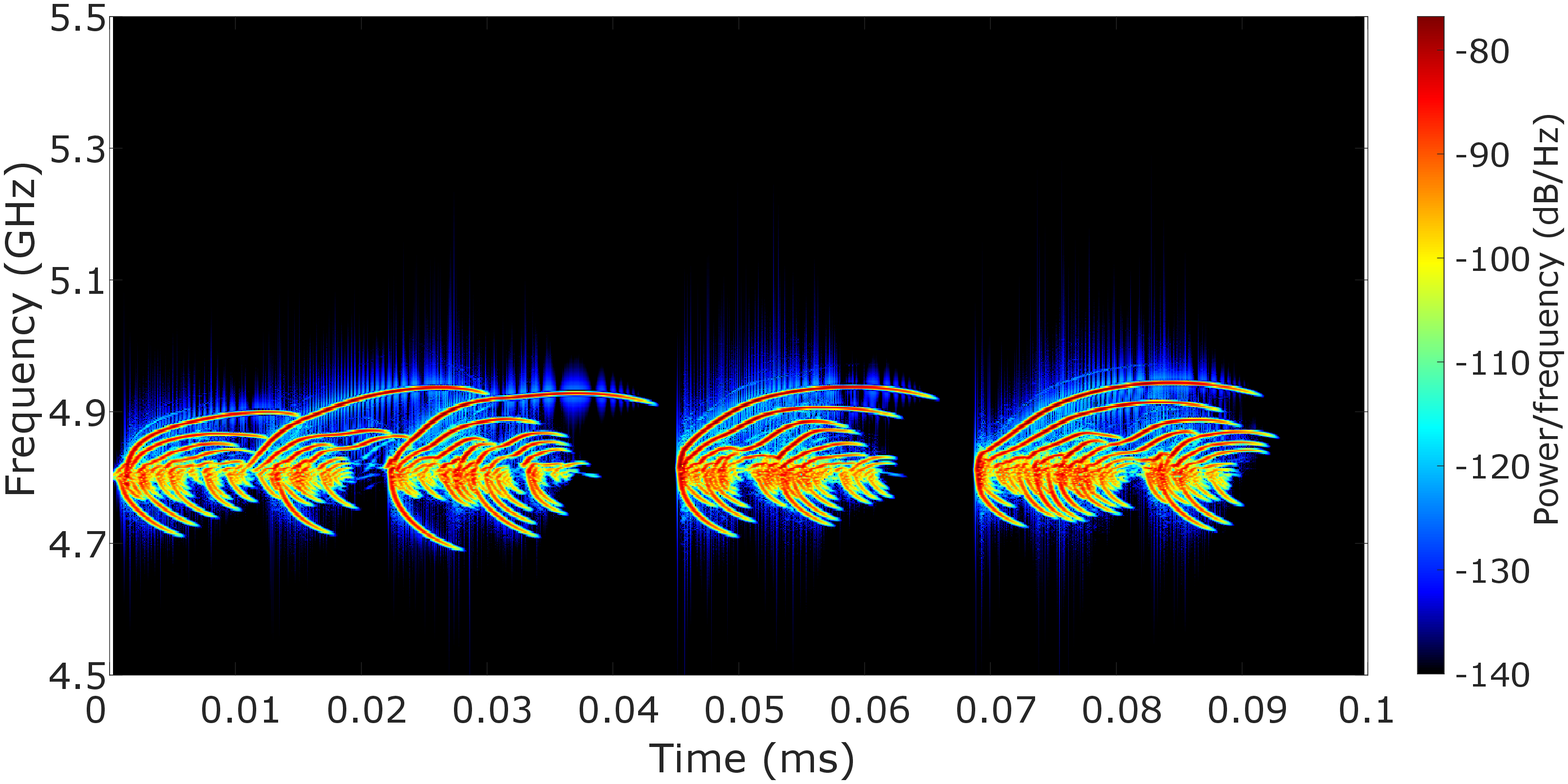}
\caption{Dynamic spectra of enhanced microwave emission during the decay stage of argon plasma: experimental data  \cite{viktorov_2016_EPL} (top) and simulation with BOT code. 
}
\label{fig3}
\end{figure}

Thus, with the simplest electrostatic model we are able to reproduce some essential features of observed signals assuming realistic physical parameters. Calculations for other frequency bands and other discharge parameters give similar results. For example, \fref{fig3} shows our very first experimental spectrogram with chirp structures and corresponding BOT simulations. This signal was detected in argon plasma (opposite to nitrogen plasma considered above). 

\section{Summary}

The Berk--Breizman  model seems to be quite suitable to explain complex transients with  periodic frequency sweeps observed in decaying plasma after high-power ECR discharge in a laboratory mirror trap. Following a traditional approach, in this paper we  consider an isolated wave-particle resonance. However, due to a naturally large difference between the bounce and cyclotron periods, we must assume a fairly large  order $n\sim30$ of the bounce resonance. At this, it is quite natural that few neighboring resonances (physically corresponding to electrons with different energies) can simultaneously drive the same unstable mode. This case, in some sense intermediate between the quasilinear and Berk--Breizman's paradigms, is still open for research. 

Although not fully realized in our experiments, there is an additional possibility to reduce the bounce resonance order exploiting the fact that $\ob/\oc$ grows inversely to the external magnetic field strength, e.g.\  $\ob/\oc\propto1/\sqrt{B}$ in our model. Thus, fast switching-off of the  magnetic field during plasma decay seems to be favorable for observation of Berk--Breizman phenomena. For example, assuming decompression factor $\a=3$ (instead of 1.25 used in our examples) would result in a shift of the resonance order to $n\sim10$ and in additional selection rules separating neighboring resonances due to a sharp dependence of  coupling parameter $V_n$ on $n$.

Another interesting observation is that frequency chirping events always start with a steady enhanced level of emission at specific fixed frequencies. Such constantly highlighted seed lines {may serve} as a possible driver for the nonlinear excitation of {\emph{subcritical} plasma instabilities. Although destabilizing role of the nonlinearity was recognized in seminal papers \cite{berk96,brei97}, the excitation of subcritical instabilities is still a matter of intriguing researches \cite{les16}}.

As a final remark, we remind that a laboratory modeling of non-stationary processes of wave-particle interactions is important for a space plasma research since there are a lot of open questions about the origin of different types of emissions in space cyclotron masers, especially mechanisms of fine spectral structure \cite{Treumann_2006}. At present, a full theory of fast frequency events in space plasma is still not finished, although the concept of electron holes in a phase-space is occasionally used, e.g.\ for interpretation of emissions in the Earth's auroral upward and downward current regions \cite{akr_holes1,akr_holes2}, Jovian S-burst emission \cite{astr15} etc. We hope that our results may advertize the application of ``holes and clumps'' paradigm to other problems of high-frequency electron dynamics.

\ack Authors are grateful to Boris Breizman for inspiring discussions and Matthew Lilley for providing the open source code BOT. This work was supported by the Russian Science Foundation (grant No.~17--72--10288).

\section*{References}
\def\etal{\textit{et al.}}

\end{document}